\documentstyle[11pt,paspconf]{article}

\begin{document}
\title{Everything you ever wanted to know about the JVAS gravitational lens system B1030+074 but were afraid to ask}
\author{E. Xanthopoulos, I. W. A. Browne, P. N. Wilkinson, N. J. Jackson, A. Karidis}
\affil{University of Manchester, NRAL Jodrell Bank, Macclesfield, Cheshire SK11 9DL, England}
\author{R. W. Porcas, A. R. Patnaik}
\affil{Max-Planck-Institut f\"{u}r Radioastromomie, Auf dem H\"{u}gel 69, D 53121, Bonn, Germany}
\author{L. V. E. Koopmans}
\affil{Kapteyn Astronomical Institute, P. O. Box 800, 9700 AV Groningen, The Netherlands}
\author{D. R. Marlow}
\affil{Department of Physics and Astronomy, University of  Pennsylvania, 209 South 33d Street, Philadelphia, PA 19104-639}

\begin{abstract}
We present an overview of all the observations (radio - VLA, MERLIN, VLBA, EVN - 
and optical - WFPC2 and NICMOS - ) that were initially used to confirm the 
gravitational lens nature of the double JVAS system B1030+074.  
Since the 1.56 arcsec system showed some first indication of variability it has
been monitored with the VLA and MERLIN to confirm its variable nature. 
We also present new VLBA observations of the lens system 
at 1.7 GHz that have unveiled detailed structure of the jet in the strong 
component and first detection of the jet in the faint component. 
\end{abstract}


\section{Introduction and data}
During the course of the Jodrell Bank VLA Astrometric Survey (JVAS) (
Browne et al. 1998, Wilkinson et al. 1998) the double gravitational
lens B1030+074 was uncovered (Xanthopoulos et al. 1998). This 1.56 arcsec separation 
double system, with a redshift of 
the lens of 0.599 and of the source of 1.535, was confirmed 
based on both radio data (VLA, MERLIN, VLBA, EVN) and optical data (HST/WFPC2 V \& I, 
NICMOS H, Keck spectra). An initial model of the system has predicted a time 
delay of 156/h$_{50}$ days.    

\begin{figure}
\plotfiddle{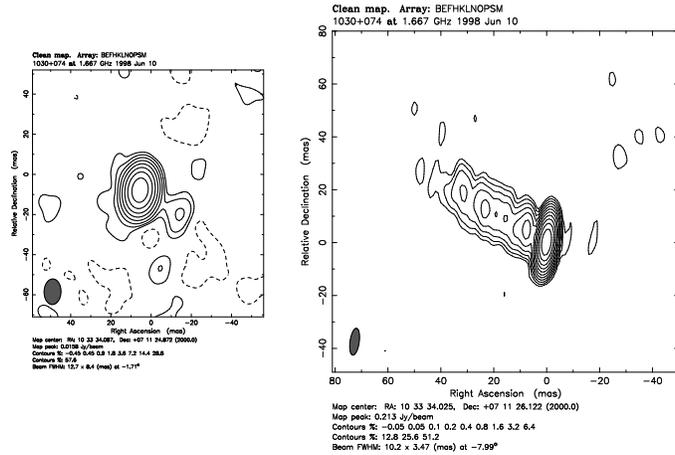}{5cm}{0}{40}{40}{-140}{-12}
\caption{The VLBA+EB 1.7 GHz maps of the two components of B1030+074; on the right 
the strong A at  
0.05\% contour loglevels of the peak value of 0.213 Jy/beam and the 
faint B component on the left at 0.45\% contour loglevels of the peak value of 0.0158 Jy/beam. } 
\end{figure}

The new VLBA+Effelsberg observations at 1.7 GHz (3mas resolution, Figure 1) 
show the jet in the A-component extending up to 60 mas and  resolved in 
3 hotspot-like features along the length of the jet.   
The corresponding jet feature in the faint B-component mirrored and with  
a hotspot feature is detected also for the first time. 
The system was monitored with the VLA from February to October 1998 
at 8.4 GHz (A, BnA and B configuration).
MERLIN long track and monitoring snapshot observations of the system were obtained in the 
C-band (5 GHz) and L-band (1.7 GHz) during 1998.  
VSOP data of B1030+074 were also obtained in May 1998 at the 1.7 GHz and 15 GHz frequencies.

\section{Conclusions}
Radio and optical observations have confirmed that B1030+074
is a gravitational lens system. 
There are 3 features that make this lens system special: a)
it is the lens system with the highest flux density ratio between its components
b) its lensing galaxy, a quasar/BL Lac early spiral type galaxy has 
   an asymmetric structure which might be part of the main body of
   the galaxy or another interacting galaxy c)
its extended jet-like structure revealed in both components from the VLBA 
   observations might make it the first lens from which a time delay 
   could be measured from differential proper motions.
B1030+074 has all the qualities of a "golden lens", a lens appropriate 
for the measurement of the Hubble constant. 

\acknowledgments
This research was supported by European Commission, TMR Programme, Research Network Contract ERBFMRXCT96-0034 \\ 
``CERES".


\begin{references}
\reference Browne, I. W. A. et al. 1998, \mnras, 293, 257 
\reference Wilkinson, P. N., et al. 1998, \mnras, 300, 790 
\reference Xanthopoulos, E. et al. 1998, \mnras, 300, 649
\end{references}
\end{document}